\documentclass[letter,longauth,tradiabstract]{aa} 
\usepackage{graphicx}
\usepackage{lscape}
\usepackage{txfonts}

\newcommand{\mic}{$\mu$m}

\newcommand{\Msun}{M$_{\odot}$}

  \def\mcfost{{\sc MCFOST}}
  \def\prodimo{{\sc ProDiMo}}

  \def\cT2{c_T^2}

\newcommand{\water}[0]{H$_2$O}

\begin{document}

\title{Gas in the protoplanetary disc of HD\,169142: {\it Herschel}'s view
\thanks{{\it Herschel} is an ESA space observatory with science instruments provided by European-led Principal Investigator consortia and with important participation from NASA. 
}}

\author{G. Meeus\inst{1}, C. Pinte\inst{2,3}, P. Woitke\inst{4,5,6}, B. Montesinos\inst{7}, I. Mendigut\'ia\inst{7}, P.~ Riviere-Marichalar\inst{7}, C. Eiroa\inst{1}, G.S.~Mathews\inst{8}, B. Vandenbussche\inst{9}, C.D. Howard\inst{10}, A. Roberge\inst{11}, G. Sandell\inst{10}, G. Duch\^ene\inst{3,12}, F. M\'enard\inst{3}, C.A.~Grady\inst{13,11}, W.R.F.~Dent\inst{14,15}, I. Kamp\inst{16}, J.C. Augereau\inst{3},  W.F. Thi\inst{5,3}, I. Tilling\inst{5}, J.M. Alacid\inst{17}, S. Andrews\inst{18}, D.R.~Ardila\inst{19}, G.~Aresu\inst{16}, D. Barrado\inst{20,7}, S. Brittain\inst{21}, D.R. Ciardi\inst{22}, W. Danchi\inst{11}, D. Fedele\inst{1,23,24}, I.~de~Gregorio-Monsalvo\inst{14,15}, A.~Heras\inst{25}, N. Huelamo\inst{7}, A. Krivov\inst{26}, J. Lebreton\inst{3}, R. Liseau\inst{27}, C. Martin-Zaidi\inst{3}, A.~Mora\inst{28}, M.~Morales-Calderon\inst{29}, H. Nomura\inst{30}, E. Pantin\inst{31}, I. Pascucci\inst{24}, N. Phillips\inst{5}, L. Podio\inst{16}, D.R.~Poelman\inst{6}, S.~Ramsay\inst{32}, B.~Riaz\inst{24}, K. Rice\inst{5}, E. Solano\inst{17}, H. Walker\inst{33}, G.J. White\inst{33,34}, J.P. Williams\inst{8}, G. Wright\inst{4}
}

\authorrunning{G. Meeus et al.}

  \institute{Dep. de F\'isica Te\'orica, Fac. de Ciencias, UAM Campus Cantoblanco, 28049 Madrid, Spain\\ 
      \email{gwendolyn.meeus@uam.es}
  \and 
 School of Physics, University of Exeter, Stocker Road, Exeter EX4 4QL, United Kingdom
  \and 
 Universit\'e Joseph Fourier - Grenoble 1/CNRS, Laboratoire d'Astrophysique de Grenoble (LAOG) UMR 5571, BP 53,  38041 Grenoble Cedex 09, France
 \and 
 UK Astronomy Technology Centre, Royal Observatory, Edinburgh, Blackford Hill, Edinburgh EH9 3HJ, UK
  \and 
 SUPA, Institute for Astronomy, University of Edinburgh, Royal Observatory Edinburgh, UK
Institute for Astronomy, University of Edinburgh, Royal Observatory, Blackford Hill, Edinburgh, EH9 3HJ, UK
 \and 
School of Physics \& Astronomy, University of St.~Andrews, North Haugh, St.~Andrews KY16 9SS, UK
 \and 
   LAEX, Depto. Astrof{\'i}sica, Centro de Astrobiolog{\'i}a (INTA-CSIC), P.O. Box 78, E-28691 Villanueva de la Ca\~nada, Spain
 \and 
 Institute for Astronomy (IfA), University of Hawaii, 2680 Woodlawn Dr., Honolulu, HI 96822, USA
 \and 
 Instituut voor Sterrenkunde, Katholieke Universiteit Leuven, Leuven, Belgium
 \and 
 SOFIA-USRA, NASA Ames Research Center, MS 211-3, Moffett Field, CA 94035, USA
 \and 
NASA Goddard Space Flight Center, Exoplanets \& Stellar Astrophysics lab, Code 667, Greenbelt, MD 20771, USA
 \and 
 Astronomy Department, University of California, Berkeley CA 94720-3411 USA
\and 
Eureka ScientiÞc, 2452 Delmer, Suite 100, Oakland, CA 96002, USA
 \and 
 ALMA, Avda Apoquindo 3846, Piso 19, Edificio Alsacia, Las Condes, Santiago, Chile
 \and 
European Southern Observatory, Alonso de C\'ordova 3107, Vitacura, Santiago, Chile
 \and 
 Kapteyn Astronomical Institute, P.O. Box 800, 9700 AV Groningen, The Netherlands
\and 
Spanish Virtual Observatory, Unidad de Archivo de Datos, Depto. Astrof\'isica, Centro de Astrobiolog\'ia (INTA-CSIC), P.O. Box 78, E-28691 Villanueva de la Ca\~nada, Spain
 \and 
Harvard-Smithsonian Center for Astrophysics, 60 Garden St., Cambridge, MA, USA
\and 
NASA Herschel Science Center, California Institute of Technology, Pasadena, USA
\and 
Calar Alto Observatory, Centro Astron\'omico Hispano-Alem\'an, c/ Jes\'us Durb\'an Rem\'on, 2-2, 04004 Almer\'ia, Spain
\and 
Department of Physics \& Astronomy, Clemson University, Clemson SC 29634-0978 USA
\and 
NASA Exoplanet Science Institute/Caltech
770 South Wilson Avenue, Mail Code: 100-22, Pasadena, CA USA 91125
\and 
Max Planck Institut f\"ur Astronomie, K\"onigstuhl 17, 69117 Heidelberg, Germany;
\and 
Department of Physics and Astronomy, Johns Hopkins University, Baltimore, MD 21218, USA
\and 
Research and Scientific Support Department of ESA, ESTEC/SRE-C, Postbus 299, 2200 AG Noordwijk, The Netherlands
\and 
Astrophysikalisches Institut und Universit{\"a}tssternwarte, Friedrich-Schiller-Universit{\"a}t, Schillerg{\"a}{\ss}chen 2-3, 07745 Jena, Germany
\and 
Department of Radio and Space Science, Chalmers University of Technology, Onsala Space Observatory, 439 92 Onsala, Sweden
\and 
ESA-ESAC Gaia SOC, P.O. Box 78, 28691 Villanueva de la Ca\~nada, Madrid, Spain
\and 
Spitzer Science Center, California Institute of Technology, 1200 E California Blvd, 91125 Pasadena, USA.
\and 
Department of Astronomy, Graduate School of Science, Kyoto University, Kyoto 606-8502, Japan
\and 
CEA/IRFU/SAP, AIM UMR 7158, 91191 Gif-sur-Yvette, France
\and 
European Southern Observatory, Karl-Schwarzschild-Strasse, 2, 85748 Garching bei M\"unchen, Germany.
\and 
The Rutherford Appleton Laboratory, Chilton, Didcot, OX11 OQL, UK
\and 
Department of Physics \& Astronomy, The Open University, Milton Keynes MK7 6AA, UK and The Rutherford Appleton Laboratory, Chilton, Didcot, OX11 OQL, UK
}

   \date{Received 2010; accepted  }
 \abstract{
In an effort to simultaneously study the gas and dust components of the disc surrounding the young Herbig Ae star HD\,169142, we present far-IR observations obtained with the PACS instrument onboard the {\it Herschel} Space Observatory. This work is part of the Open Time Key Project GASPS, which is aimed at studying the evolution of  protoplanetary discs. To constrain the gas properties in the outer disc, we observed the star at several key gas-lines, including [OI] 63.2 and 145.5\,\mic, [CII] 157.7\,\mic, CO 72.8 and 90.2\,\mic, and o-\water \ 78.7 and 179.5\,\mic. We only detect the [OI] 63.2\,\mic\,line in our spectra, and derive upper limits for the other lines. We complement our data set with PACS photometry and $^{12/13}$CO data obtained with the Submillimeter Array. Furthermore, we derive accurate stellar parameters from optical spectra and UV to mm photometry. We model the dust continuum with the 3D radiative transfer code \mcfost \ and use this model as an input to analyse the gas lines with the thermo-chemical code \prodimo. Our dataset is consistent with a simple model in which the gas and dust are well-mixed in a disc with a continuous structure between 20 and 200 AU, but this is not a unique solution. Our modelling effort allows us to constrain the gas-to-dust mass ratio as well as the relative abundance of the PAHs in the disc by simultaneously fitting the lines of several species that originate in different regions. Our results are inconsistent with a gas-poor disc with a large UV excess; a gas mass of 5.0 $\pm \ 2.0\, \times$ 10$^{-3}$ \Msun\, is still present in this disc, in agreement with earlier CO observations. 
}
 
  \keywords{Stars: pre-main sequence; (Stars:) planetary systems; (Stars:) circumstellar matter; protoplanetary discs;  Infrared: planetary systems}
 
 \maketitle
 

\section{Introduction}

Giant gas-planets form in protoplanetary discs within the first 10 Myr after protostar formation. Therefore the amount of gas that is present in a disc at a given time is very important, as it determines whether these planets can still be formed. The study of the dust is equally important, as it witnesses the first steps of planet formation. A lot is known about the dust in protoplanetary discs thanks to the Space Observatories ISO and {\it Spitzer}\,: different degrees of dust processing were observed  with no clear relation to stellar properties, while grain growth could be related to the disc structure (e.g. Sicilia-Aguilar et al. \cite{sicilia2007}, Meeus et al. \cite{meeus2009}). Gas is more difficult to observe, as the spectral lines are not very strong. Most gas studies are based on CO lines in the near-IR (warm gas) or in the mm, where CO freeze-out onto grains is a complicating factor (e.g. Brittain et al. \cite{brittain2007}, Dent et al. \cite{dent2005}). However, as gas lines are stronger in the far-IR, we expect this field to experience major breakthroughs in the coming years based on data obtained with the {\it Herschel} Space Observatory (Pilbratt et al. \cite{pilbratt2010}), which provides sensitive far-IR photometry and spectroscopy. 

The source HD 169142 is a young - age 6$^{+6}_{-3}$ Myr (Grady et al. \cite{grady2007}) - Herbig Ae star with an IR to millimetre excess attributed to a circumstellar (CS) disc. Submillimeter Array (SMA) observations show a disc in Keplerian rotation with radius $r$ = 235 AU and inclination $i \sim$ 13\degr  (Raman et al. \cite{raman2006}) and a total gas mass of 0.6-3.0 $\times 10^{-2}$ \Msun \ (Pani\'c et al. \cite{panic2008}). For a Herbig Ae, the star is unusual as it has a small near-IR excess (e.g. Dominik et al. \cite{dominik2003}). Furthermore, SWS/ISO spectra revealed that the silicate 10\,\mic \ feature, detected  in the majority of the Herbig Ae/Be stars (only 8 out of 53 lack the feature, Juh\'asz et al., \cite{juhasz2010}) is lacking in HD\,169142 (Meeus et al. \cite{meeus2001}). This absence can be explained if the silicate grains are either too large or too cold to emit at 10\,\mic \ (Meeus et al. \cite{meeus2002}). On the other hand, features of polycyclic aromatic hydrocarbons (PAHs), which can be excited by UV photons, were clearly detected with ISO and {\it Spitzer}. Based on the near to far-IR ratio, Grady et al. (\cite{grady2007}) suggested that the inner region has already cleared some material, and that the inner and outer disc are not coupled.  However, this might also be explained by a reduced opacity due to grain growth. 

In this paper we show the first {\it Herschel} observations of HD\,169142, and use radiative transfer and chemical models to constrain the gas properties of the {\em outer} disc.


\section{{\it Herschel} PACS observations}
\label{s_pacs}

\begin{figure}
\begin{center}
\includegraphics[width=6.5cm,angle=90]{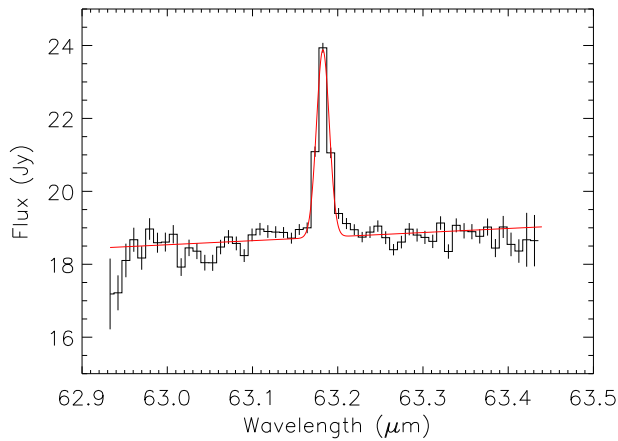}
  \caption{Observed [OI] 63\,\mic \ emission line (black) together with our line fit (red; parameters listed in Table~\ref{t_lines}).   }
    \label{f_OI_line}
    \vspace{-5mm}
 \end{center}
\end{figure}

Our observations are part of the {\it Herschel} Open Time Key Project GASPS (P.I. Dent, see Mathews and Thi, this issue, 
and Dent et al. \cite{dent2010}).
We obtained PACS (Poglitsch et al. \cite{poglitsch2010}) photometry (obsid 1342183656, Point Source, 400\,s) and spectroscopy (obsid 1342186309, PacsRangeSpec, 5150\,s and obsid 1342186310, PacsLineSpec, 1669\,s). The spectroscopic data were reduced with the developer build version 3.0.1212 of the {\it Herschel} Interactive Processing Environment (HIPE; Ott \cite{ott2010}), using standard tasks provided in HIPE. In order to conserve the best signal and not to introduce additional noise, we only extracted the central spaxel and corrected for the flux loss with an aperture correction. Furthermore, we applied another correction factor as supplied by the PACS team (1.3 in the blue, and 1.1 in the red), to obtain a flux calibration accuracy of 40\%. More details on the data reduction can be found in Mathews et al. (\cite{mathews2010}). In Fig.\,\ref{f_OI_line} we show our only detection, the fine structure line [OI] 63.2\,\mic. We also searched for other gas lines that were targeted in our spectroscopy, but did not detect:  [OI] 145.5\,\mic, [CII] 157.7\,\mic, CO 72.8 and 90.2\,\mic, nor o-\water \ 78.7 and 179.5\,\mic. In Table~\ref{t_lines} we list the measured line properties. We note that in the LWS/ISO spectrum of HD\,169142, the [CII] 157.7\,\mic \ line was detected, but LWS uses a larger beam ($\sim$ 40\arcsec) than the size of the central spaxel of PACS (9\farcs4 by 9\farcs4). Therefore we also analysed the other 24 spaxels of the IFU and found a tentative detection in spaxels 13 and 14, which are to the West of HD\,169142. The cause of this emission (if confirmed) is currently unknown.

 \begin{table}[t]
  \caption[]{Summary of the PACS targeted lines and the CO lines that were observed with the SMA (Raman et al. \cite{raman2006}, Pani\'c et al. \cite{panic2008}). We list the central wavelength, continuum and line flux or upper limits (three sigma) in the case of a non-detection. Between brackets we give the one sigma statistical error for the detections. 
  }
  \begin{tabular}{rcccc}
    \hline\hline 
Species       &                       Transition                         & $\lambda_{c}$   & Continuum     & Line Flux\\
                      &                                                                 & (\mic)                   &      (Jy)              & (10$^{-18}$ W/m$^{2}$)\\
\hline   
[OI]                 & $^3P_1 \rightarrow \ ^3P_2$            & 63.18                 & 18.89 (0.13)     &71.7 (3.8)\\

[OI]                 & $^3P_0 \rightarrow \  ^3P_1$            & 145.53               & 13.76 (0.03)   &$<10.4$ \\  

[CII]                & $^2P_{3/2} \rightarrow \ ^2P_{1/2}$ & 157.74               & 14.48 (0.03)   &$<6.4$   \\   
o-\water   \    & 2$_{12} \rightarrow 1_{01}$             & 179.53               & 13.12 (0.05)     &$<8.8$   \\  
o-\water   \    & 4$_{23} \rightarrow 3_{12}$             & 78.74                 &  17.82 (0.32)    &$<10.6$ \\  
CO                 & 36 $\rightarrow$ 35                            & 72.84                &  17.92 (0.04)    &$<15.5$ \\  
CO                 & 29 $\rightarrow$ 28                            & 90.16                &  17.00 (0.05)    &$<10.6$ \\  
\hline
  $^{12}$CO & 2 $\rightarrow$ 1                                & 1300.40             & 0.169 (0.005)   &  0.093 (0.004)\\
  $^{13}$CO & 2 $\rightarrow$ 1                                & 1360.22             & 0.169 (0.005)   &  0.048 (0.004)\\
\hline
\end{tabular}
\label{t_lines}
\end{table}

Photometric observations were obtained in Point Source Mode for both the blue (70\,\mic) and red (160\,\mic) bands, and  reduced using HIPE v2.3.4 with the Point Source Observations pipeline. The source is spatially unresolved, with FWHM 5\arcsec \ at 70\,\mic \ and 11\arcsec \ at 160\,\mic. We derived photometry using an aperture of 21\arcsec, and applied an aperture correction from the PACS PhotChopNod Release Note (Feb. 22, 2010). This gives a flux of 27.35 $\pm$ 0.03 Jy at 70\,\mic \ and 17.39 $\pm$ 0.05 Jy at 160\,\mic, with a flux calibration uncertainty of 5\% in the blue, and 10\% in the red. Applying the flux uncertainty of 40\% for the spectroscopy, our photometry is consistent with the continuum fluxes measured from the spectral scans. Given the smaller error in the photometry, however, we give preference to those fluxes, and expect to reduce the spectroscopic flux uncertainties in the future.

\section{Determination of stellar parameters}
\label{s_para}

Because previous studies list conflicting effective temperatures, we re-estimated the fundamental parameters. Photometry excluding the contribution from the cool companion (at a distance of 9\farcs3; Grady et al. \cite{grady2007}) was collected from Cutri et al. (\cite{cutri2003}), Zacharias et al. (\cite{zacharias2004}) and Sylvester et al. (\cite{sylvester1996}). We also analysed optical spectra, taken with CES/CAT with a resolution of $\sim$ 60000. Temperatures were selected by simultaneously fitting the lines from neutral {\it and} singly ionised species. The synthetic stellar spectra were computed using the ATLAS9 and SYNTHE codes by Kurucz (\cite{kurucz1993}).  $\chi^2$ tests show that models between $T_{\rm eff}\!=\!7500$\,K, $\log g_*\!=\!4.0$, [Fe/H]=$-0.50$ and $T_{\rm eff}\!=\!7800$\,K, $\log g_*\!=\!4.1$, [Fe/H]=$-0.25$ match our data. 

Furthermore, the International Ultraviolet Explorer (IUE) obtained five spectra of HD\,169142: one between 1200--1900\,\AA{} (SW) and four between 1900--3200\,\AA{} (LW). The SW spectrum is unusable for measurements below 1650\,\AA, but suggests emission lines of OI (1304\,\AA), CII (1335\,\AA) and CIV (1550\,\AA) - unfortunately, most are dominated by bad pixels. The LW spectra do not show any variability (over a period of 4.5 months),  match the stellar photosphere, and do not show emission features. Overall, the IUE data  (obtained from the INES INTA archive) do not give evidence for a UV excess.

\section{Analysis}
\label{s_ana}

\subsection{Dust and continuum modelling}

Far-IR lines emerging from a CS disc are affected by the stellar UV irradiation, disc mass and geometry, dust size and composition, as well as PAH abundance.  In order to interpret the lines observed with {\it Herschel}, it is crucial to first obtain a solid knowledge of the dust structure in the disc, based on as many observations as possible, like SED, scattered light images and visibilities.  Each of these observations provide complementary views of the disc structure and the dust properties. The disc model is calculated with the Monte Carlo radiative transfer code \mcfost\ (Pinte et al. \cite{pinte2006,pinte2009}), as outlined below. 

We consider an axisymmetric, slightly flared density structure with a Gaussian vertical profile, assuming power-laws for the surface density and scale height. We assume homogeneous and spherical dust grains (Mie theory), with sizes distributed according to the power-law $f(a) \propto a^{-3.5}$ between $a_{\mathrm{min}}$ and $a_{\mathrm{max}}$, where $a$ is the grain radius.  The dust is assumed to be well-mixed with the gas, i.e. the dust/gas ratio is constant throughout the disc. The star is reproduced by a uniformly radiating sphere with previously determined parameters $T_{\rm eff}\!=\!7800$\,K, $\log g_*\!=\!4.1$, [Fe/H] = -0.25, and R$_{*}$ = 1.6\,R$_\odot$. Parameters are adjusted to simultaneously fit the SED, the \emph{Spitzer}/IRS spectrum (Sloan et al. \cite{sloan2005}), the $1.1\,\mu$m HST image (Grady et al. \cite{grady2007}) and the 1.3\,mm SMA visibilities (Pani\'c et al. \cite{panic2008}). The scattered light image mainly constrains the flaring index to a low value (around 1.0) and the mm visibilities indicate a surface density varying as $r^{-1}$. 

We reproduce the 1.3\,mm emission with a dust mass of $1.5\times 10^{-4}$\,\Msun \ with a grain size distribution\footnote{Note that for these size distribution parameters, only 31.5\,\% of the dust mass is in grains $\leq\!1\,$mm, the rest (68.5\%) is in larger grains.} between $a_{\rm min}\!=\!0.03\,\mu$m and $a_{\rm max}\!=\!1\,$cm. The weak emission around 10\,$\mu$m implies that there is a discontinuity in the disc $\tau\!=\!1$ surface, possibly due to a shadowed area or a gap. In this paper, we explore a solution with a gap. For the dust composition, we use a mixture of 70\,\% silicates (Draine et al. \cite{draine2003}) and 30\,\% amorphous carbon (ACAR sample; Zubko et al. \cite{zubko1996}), and calculate the effective optical index with the Bruggeman mixing rule. The strong PAH features around 10\,$\mu$m can be reproduced with a low PAH abundance due to the low continuum emission in that region. We modelled the emission with a single grain-size (54 carbon atoms, positively ionised) and assume that the PAHs are uniformly distributed in the outer disc. We did not try to reproduce in detail the various bands observed with IRS, but rather constrain their abundance. We obtain M$_{\rm PAH}$/M$_{\rm dust}$ = $5\times 10^{-4}$, corresponding to $f_{\rm PAH}\!=\!0.03$ (where $f_{\rm PAH}$ is the uniform PAH abundance relative to a standard ISM abundance of $10^{-6.52}$ PAH particles/H-nucleus, $m_{\rm PAH}\!=\!667\,$amu, and gas/dust = 100).

\begin{figure}
\begin{center}
 \includegraphics[width=8.3cm]{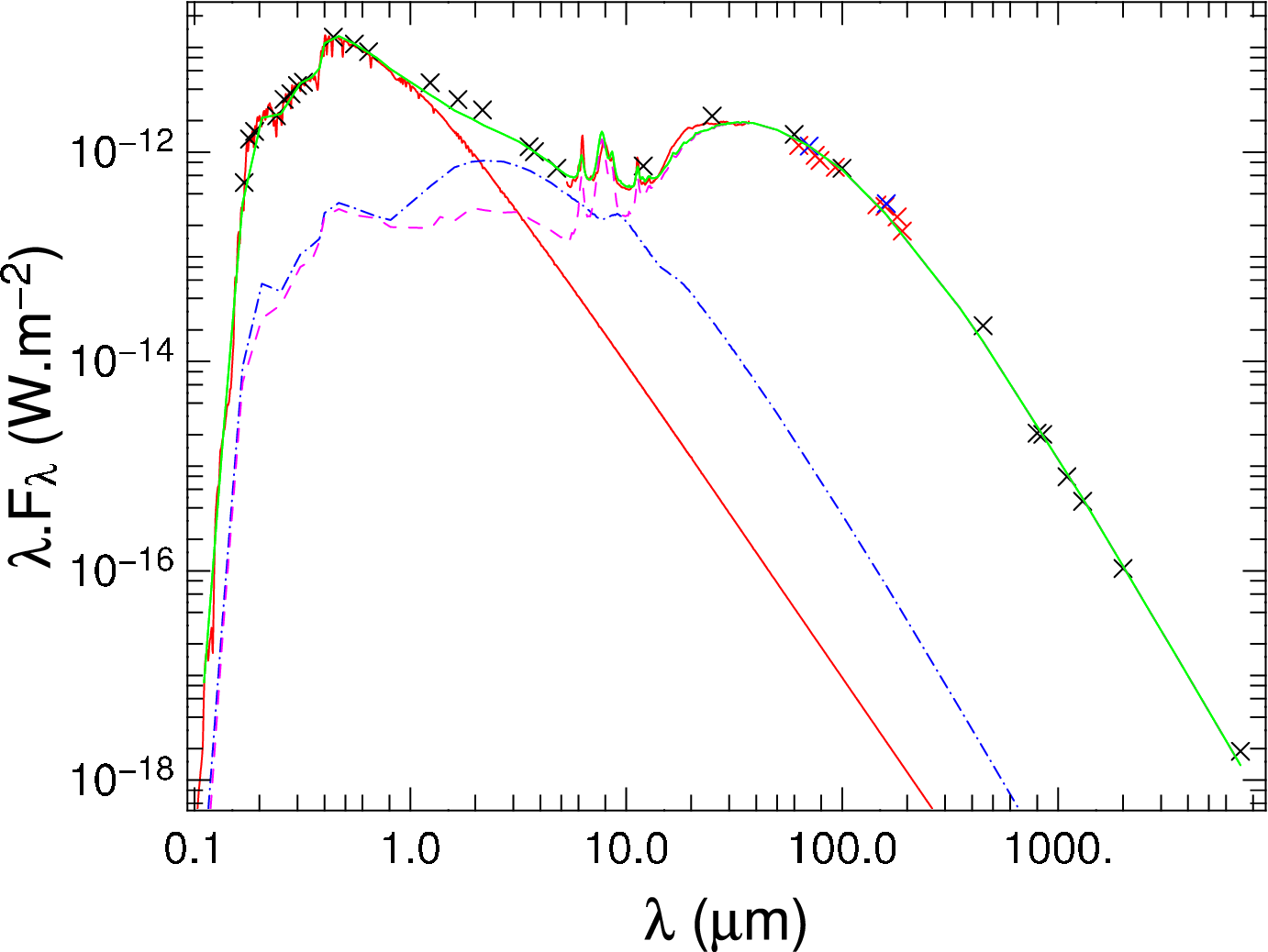}
  \caption{Best fit with \mcfost\, to the observed data. Shown is the photometry obtained from the literature
    (black), {\it Spitzer} IRS spectrum (red), PACS photometric observations (blue crosses), and PACS continua derived from the spectroscopic observations (red crosses). The green line represents the best \mcfost\ model, in red the atmosphere model. The blue dot-dashed and pink dashed lines are the contributions (scattered light and thermal emission) from the inner and outer disc, respectively.}
   \label{fig:SED}
    \vspace{-5mm}
\end{center}
 \end{figure}

\begin{table}[htbp]
 \caption{Parameters of the best fitting dust model. 
 \label{tab:best_parameters}} 
 \centering
  \begin{tabular}{lll}
    \hline
    \hline
    Parameter & Inner Disc & Outer Disc\\
    \hline
    $r_\mathrm{in}$ (AU)& 0.1 & 20 \\
    $r_\mathrm{out}$ (AU)& 5 & 235 \\
    surface dens. exp $\epsilon$ &  -1.0 & -1.0\\
    flaring exponent $\beta$ & 1.05 & 1.00\\
    ref.~scale height $h_0$ & 0.07\,AU @ 1\,AU & 12.5\,AU @ 100\,AU\\
    M$_\mathrm{dust}$ (\Msun)& $2\times 10^{-9}$ & $1.5\times 10^{-4}$\\
    $f_{\rm PAH}$ & 0 & 0.03\\
    \hline
  \end{tabular}
 \end{table}

In Fig.\,\ref{fig:SED} we show the best fitting model on top of the SED. Our model provides good constraints on the outer disc (mid-, far-IR and mm emission): dust mass, scale height, flaring index and surface density profile, as well as dust properties and amount of PAHs. However, the inner disc (near-IR emission) remains poorly constrained. We tried to improve the fit by moving the inner radius closer to the star ($r_{\rm in}=0.06$\,AU), but then the temperature is too high (2400\,K) for dust, even carbon, to survive. Because we lack simultaneous visible and near-IR photometry as well as spatially resolved near-IR data, we did not further improve the fit to the inner disc. However, as most of the mid-IR and mm lines originate from the outer disc, this is not critical for the present paper.

\subsection{Gas and line modelling}

\begin{table}
  \caption{Predicted line fluxes for different models, compared with the observed fluxes. The best fit is obtained by model\,\#3. All models in this table have turbulent line broadening $v_{\rm turb}$ = 0.15\,km/s.}
  \begin{tabular}{lcccc}
    \hline
    \hline
                                     & model \#1   & model \#2  & model \#3 & observed \\
   gas/dust                   & 1                  & 100              & 33             & -- \\ 
   $f_{\rm PAH}$         & 0.01            & 0.0055         & 0.0087     & -- \\
   $f_{\rm UV}$            & 0.005          & 0.0                & 0.0             &  --  \\
    \hline
    & & & &\\[-2.2ex]
    Line             &\multicolumn{4}{c}{Line Fluxes $\rm[10^{-18}W/m^2]$} \\
    \hline
    [OI] 63.2\,\mic         & 154        & 71.6       & 71.6       & 71.7 \\
                                         
    [OI] 145.5 \,\mic      & 5.17       & 10.1       & 7.01       & $<$\,10.4 \\
                                         
    [CII] 157.7\,\mic      & 4.58       & 0.04      & 0.06      & $<$\,6.4 \\
o-\water\,179.5\,\mic & 5.66      & 5.15       & 1.76       & $<$\,8.8 \\  
                                         
\hline                                   
    $^{12}$CO 2 $\rightarrow$ 1             
                            & 0.060     & 0.092     & 0.093     & 0.093 \\
    $^{13}$CO 2 $\rightarrow$ 1             
                            & 0.011     & 0.059     & 0.048     & 0.048 \\
    $^{12}$CO$/^{13}$CO                    
                            & 5.69       & 1.55       & 1.92       & 1.94  \\
  \hline
  \end{tabular}
  \label{tab:lines}
\end{table}

\begin{figure}
 \includegraphics[width=8.6cm]{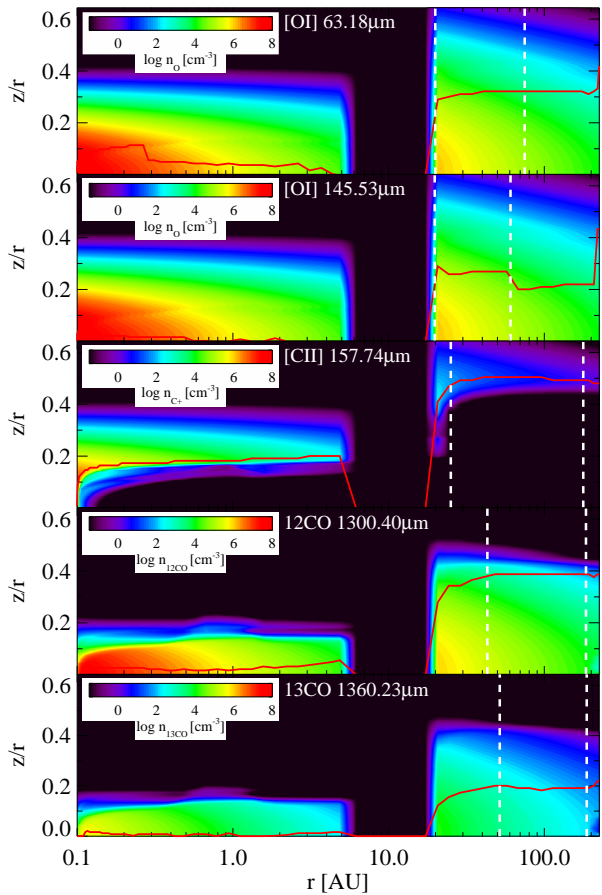}\\
    \vspace{-5mm}
  \caption{Spatial origin of the various gas emission-lines.  The two vertical white dashed lines indicate 15\% and 85\% of the radially cumulative face-on line flux, respectively, i.e.\ 70\% of the line flux originate from within the two white dashed lines. The red lines mark the cells that contribute most to the line flux in their vertical column.}
  \label{LineAnalysis}
\end{figure}

The best model reproducing the continuum observations is now fed into the gas thermo-chemical code \prodimo\, (Woitke et al. \cite{woitke2009}) to calculate the chemical and gas temperature structure in the disc and to predict the line fluxes, following the pipeline described in Woitke et al. (\cite{woitke2010}; see their Fig.\,1). This final modelling step has additional free parameters such as the dust/gas ratio. We computed models for the following parameters: gas/dust mass ratio $\in\![1,100]$, turbulent broadening with $v_{\rm{turb}}\!\in\![0,0.15]\,$km/s, PAH abundance $f_{\rm   PAH}\!\in\![0,0.06]$ and stellar UV excess $f_{\rm UV}\!=\!L_{\rm   UV}/L_\star\!\in\![0,0.005]$ (for details, see Woitke et al. \cite{woitke2010}). Due to the stellar UV irradiation, the models generally result in disc surface-layers where the gas is warmer than the dust. Table~\ref{tab:lines} shows the calculated line fluxes for a few selected models. An important result from our modelling effort is that the gas heating by PAHs plays a central role for the line flux predictions, in our case useful for the temperature sensitive $\rm[OI]\,63.2\,\mu$m line. 
For each selection of the parameters dust/gas and $f_{\rm UV}$, we tuned the PAH abundance until a fit with the observed $\rm[OI]\,63.2\,\mu$m line flux was obtained, if possible. However, models with UV excess $f_{\rm UV}\!=\!0.005$ result in a much too hot gas and hence too strong gas emission lines, even for $f_{\rm PAH}\!=\!0$, and even if we decrease gas/dust$\,\to\!1$, as in model\,\#1.
For the standard ratio $\rm gas/dust\!=\!100$, we found model\,\#2, which fits the $\rm[OI]\,63.2\,\mu$m and $^{12}$CO line fluxes, but the $^{13}$CO line is slightly too high (2.5 sigma). A better fit is obtained with $\rm gas/dust\!=\!33$ (model\,\#3),  which simultaneously fits all three detected lines, and agrees with all other line upper limits. 
However, we emphasise that apart from the uncertainties in the observations, systematic uncertainties in the physical description of various chemical and radiative processes in the models render a proper gas mass determination difficult. The observational uncertainty in the $^{13}$CO line flux - which appears to be a dominant gas mass tracer - translates to a ratio range $\rm gas/dust\!=22$ to 50. An increase of the [OI]\,63.2\,\mic \ flux by 40\% (to account for the calibration uncertainty and match the photometry) would indicate a higher gas temperature. In our modelling, this is best interpreted by a larger PAH abundance ($f_{\rm PAH}$ = 0.02), as the other more relevant parameter, $f_{\rm UV}$, is constrained by the non-detection of the [CII]\,157.7\,\mic \ line.

The spatial distribution of the $^{12/13}$CO molecules predicted by our chemical model is consistent with the interferometric observations by Pani\'c et al. (\cite{panic2008}). Figure~\ref{LineAnalysis} shows a detailed analysis of the spatial origin of the various emission lines, as derived from  model\,\#3. The [OI]\,63.2\,$\mu$m and 145.5\,$\mu$m lines originate from a radial disc region extending from the beginning of the outer disc at 20\,AU to about 75\,AU and 60\,AU, respectively. Both lines are optically thick ($\tau_{\rm line}\!\approx 10-100$) and come from relative heights $z/r\!\!\approx\!0.2-0.3$.  The [CII]\,157.7\,$\mu$m line is optically thin and extraordinarily weak in this model, due to the lack of stellar UV photons $<\!110\,$nm that are capable of ionising carbon ($f_{\rm UV}\!=\!0$).  The $^{12}$CO and $^{13}$CO $J\!=2\!\to\!1$ lines probe the conditions in the outermost disc regions $43-185$\,AU and $52-190$\,AU, respectively, and are extremely optically thick ($\tau_{\rm line}\!\approx 100-5000$). Whereas the $^{12}$CO line comes from the PDR-like CO surface $z/r\!\approx\!0.4$, the more transparent $^{13}$CO line probes slightly deeper layers $z/r\!\approx\!0.2$. These findings are quite stable and robust among all calculated models. 

\section{Conclusion}
\label{s_discu}

We presented the first PACS observations of HD\,169142 and showed the unique capability of {\it Herschel} to obtain an independent gas mass determination by simultaneously modelling the atomic fine-structure lines that are temperature sensitive (due to their high excitation energies), when combined with ground-based observations of $^{12/13}$CO-lines. We showed that the observations are consistent with a simple model of a disc hosting a gap, with a continuous structure between 20 and 200 AU in which the gas and dust are well-mixed, but stress that this solution is not unique. We determined the location of the emitting species, and constrained the gas/dust ratio to be $\sim$ 22-50. We also derived that the UV excess emission, if present, must be lower than $f_{\rm UV}$ = 0.005. We derive a gas mass between 3.0 and 7.0 $\times$ 10$^{-3}$ \Msun, in agreement with the gas mass determination based on the CO lines alone (Pani\'c et al. \cite{panic2008}). Despite indications that the HD\,169142 disc is transitional, it still is a gas-rich disc. Furthermore, although there is dusty material close to the central star, the UV excess induced by accretion is extremely weak, if at all present. We will apply a similar modelling strategy for future GASPS observations, with the aim to study gas dissipation in a wide range of protoplanetary discs.

\begin{acknowledgement}

We thank the PACS instrument team for their dedicated support, O. Pani\'c for a discussion on her data and A. Juh\'asz for providing the {\it Spitzer}/IRS data. 
C. Eiroa, G. Meeus, I. Mendigut\'ia and B. Montesinos are partly supported by AYA 2008-01727; J.M. Alacid and E. Solano by AYA 2008-02156 and I. d. Gregorio by AYA 2008-06189 (Spanish grants). C. Pinte acknowledges funding from the EC 7$^{th}$ FP as a Marie Curie Intra-European Fellow (PIEF-GA-2008-220891). 
J.-C. Augereau, G. Duch\^ene, J. Lebreton, C. Martin-Zaidi, F. M\'enard and C. Pinte acknowledge PNPS, CNES and ANR (contract ANR-07-BLAN-0221) for financial support. D.R. Ardila, S.D. Brittain, B. Danchi, C.A. Grady, I. Pascucci, B. Riaz, A. Roberge, G. Sandell and C.D. Howards acknowledge NASA/JPL for funding support. 
PACS has been developed by a consortium of institutes led by MPE (Germany) and including UVIE (Austria); 
KUL, CSL, IMEC (Belgium); CEA,  OAMP (France); MPIA (Germany); IFSI, OAP/AOT, OAA/CAISMI, LENS, SISSA
(Italy); IAC (Spain). This development has been supported by the funding agencies BMVIT (Austria), ESA-PRODEX 
(Belgium), CEA/CNES (France), DLR (Germany), ASI (Italy), and CICT/MCT (Spain).

\end{acknowledgement}


{}

\end{document}